\documentclass[letterpaper,journal]{IEEEtran}

\usepackage{amsmath,amssymb,amsfonts}
\usepackage{graphicx}
\usepackage{cite}
\usepackage{algorithm}
\usepackage{algorithmic}
\usepackage{array}
\usepackage{booktabs}
\usepackage{balance}
\usepackage{url}
\usepackage{epstopdf}
\usepackage[caption=false,font=footnotesize,labelfont=sf,textfont=sf]{subfig}

\begin{document}

\title{CMT-Aware Channel Modeling and Transmit-Power Minimization for Pinching-Antenna Systems}

\author{Chuang~Luo,~\IEEEmembership{Graduate Student Member,~IEEE,}
        Gui~Zhou,~\IEEEmembership{Member,~IEEE,}
        Vasilis~K.~Papanikolaou~\IEEEmembership{Member,~IEEE,}
        Yingzhuang~Liu,
        and~Robert~Caiming~Qiu~\IEEEmembership{Fellow,~IEEE}\vspace{-2.5em}

\thanks{Chuang~Luo, Gui~Zhou, Yingzhuang~Liu, and Robert~Caiming~Qiu are with the School of Electronic Information and Communications (EIC), Huazhong University of Science and Technology (HUST), Wuhan 430074, China (e-mail: \{chuang\_luo, gui\_zhou, liuyz, caiming\}@hust.edu.cn). Gui Zhou is the corresponding author.

Vasilis~K.~Papanikolaou is with the Institute for Digital Communications, Friedrich-Alexander Universit{\"a}t of Erlangen-N{\"u}remberg (FAU), 91054 Erlangen, Germany (e-mail: vasilis.papanikolaou@fau.de).
}
}

\maketitle

\begin{abstract}
This letter investigates transmit-power minimization for multiuser pinching-antenna system (PAS) from a coupled-mode-theory (CMT)-aware perspective. Existing CMT-based pinching antenna (PA) studies reveal coupling-induced power exchange and radiation behavior, but these effects have not been fully embedded into system-level multi-PA channel modeling and beamforming design. We therefore develop a directional and loss-aware channel model that captures coupling-length-dependent power extraction and the downstream guided-power reduction caused by in-waveguide attenuation and upstream extraction. The model shows that PA design should account for both directional radiation and guided-power evolution, rather than only propagation distance or maximum coupling considered in most existing works. Based on this channel model, we formulate a quality-of-service (QoS)-constrained power minimization problem for continuous PA positioning and finite-codebook activation. For each candidate coupling length, the element-wise positioning and BPSO-based activation use a closed-form zero-forcing (ZF) power metric for low-complexity configuration ranking, thereby avoiding repeated beamforming optimization while excluding rank-deficient candidates and ordering the remaining ones. The selected configuration for each coupling length is then evaluated by optimal fixed-configuration QoS beamforming. Simulation results demonstrate that CMT-aware modeling fundamentally reshapes the preferred PA configuration, maximum coupling is not always power-efficient due to suppressed downstream PA contributions, and finite-codebook activation combined with ZF-based ranking provides a balance between transmit-power performance and deployment complexity.

\end{abstract}


\vspace{-1em}\section{Introduction}
\IEEEPARstart{P}{inching}-antenna systems (PAS) have recently emerged as a promising antenna architecture for establishing flexible line-of-sight (LoS) links in dense wireless environments. By guiding radio-frequency signals through low-loss dielectric waveguides and radiating them into free space through small dielectric pinching antennas (PAs), PAS can create reconfigurable radiating points along the waveguides. This capability is particularly attractive for dense indoor, industrial, and hotspot deployments, where short-range LoS links are often more reliable and power-efficient than long-distance free-space transmission from fixed antenna arrays \cite{ding2025flexible,liu2026pass}.

The performance of PAS depends critically on how the guided signal is coupled out of the waveguide and converted into free-space radiation. Many existing system-level studies adopt simplified PA models, such as isotropic point radiation or prescribed power allocation among activated PAs, to achieve tractable channel and beamforming designs. To improve the physical consistency of PAS modeling, coupled-mode theory (CMT) has been introduced by modeling a PA as an open-ended directional coupler \cite{wang2025modeling}. This model characterizes coupling-induced power exchange between the main waveguide and the pinched element, and motivates equal-power or proportional-power radiation models for transmit and pinching beamforming. Following this line, adjustable power-radiation models exploit the waveguide-PA spacing to control the radiated power ratio, with emphasis on power allocation and discrete activation optimization \cite{xu2026power}. These works provide important CMT-based foundations. However, CMT is mainly used to quantify extracted or radiated power, with the associated radiation pattern, directional gain, and coupling-length-dependent directivity not yet fully incorporated into system-level PA placement and beamforming optimization.

This limitation becomes more pronounced in multi-PA waveguide-fed PAS, where activated PAs on the same waveguide are coupled through a common guided signal. Hence, each PA contribution depends on its radiation directivity and the available guided power, which is reduced by in-waveguide attenuation and power extraction by upstream activated PAs. Recent physics-aware PAS studies have examined these effects from different aspects. Specifically, a CMT-based closed-form radiation-pattern model in \cite{zubair2026closed} provides analytical characterization of coupling-length-dependent PA directivity, offering valuable physical insights into PAS radiation. Wideband and attenuation-aware PAS models characterize antenna coupling and in-waveguide loss \cite{xiao2025ofdm,xu2025attenuation}, while the directional and pattern-aware frameworks investigate practical PA patterns for placement design \cite{zhang2025directional,feng2026measured}. These CMT-related propagation and radiation effects jointly determine the effective channel seen by users and therefore fundamentally influence system-level resource allocation and PA configuration. Ignoring them may result in inaccurate performance evaluation and suboptimal design decisions. This calls for a unified CMT-aware system-level channel model that explicitly incorporates practical electromagnetic behaviors into multi-PA PAS optimization.

Based on this CMT-aware system-level modeling framework, this letter studies quality-of-service (QoS)-constrained transmit-power minimization for multiuser PAS, aiming to clarify how CMT-induced radiation affects PA positioning, activation, and beamforming. Building upon the closed-form in-plane CMT radiation-pattern model in \cite{zubair2026closed}, we incorporate its coupling-length-dependent directional response into a multi-PA downlink channel, yielding a unified CMT-aware model that captures coupling-dependent power extraction, directional radiation, guided-power depletion, and in-waveguide attenuation. The resulting design is highly nonconvex because the PA configuration determines both the spatial channel and guided-power evolution, as shown in Fig.~\ref{fig:system_model}(c). We therefore develop two complementary methods: an element-wise continuous-positioning scheme that exploits flexible PA placement and a binary particle swarm optimization (BPSO)-based candidate-grid activation scheme for practical implementation. A closed-form zero-forcing (ZF) transmit-power metric derived from the same channel enables low-complexity ranking of candidate PA configurations according to the power required to meet the prescribed QoS targets, while excluding rank-deficient candidates. The top-ranked configuration for each coupling length is then evaluated by optimal fixed-configuration beamforming. Simulation results verify that CMT-aware modeling reshapes the preferred PA configuration and that finite-codebook activation provides a practical performance-complexity tradeoff.

\vspace{-1em}\section{System Model}
\vspace{-0.2em}\subsection{System Geometry and Unified PA Position Model}\label{System_Geometry}
\begin{figure}[!t]
\centering
\includegraphics[width=3.5in]{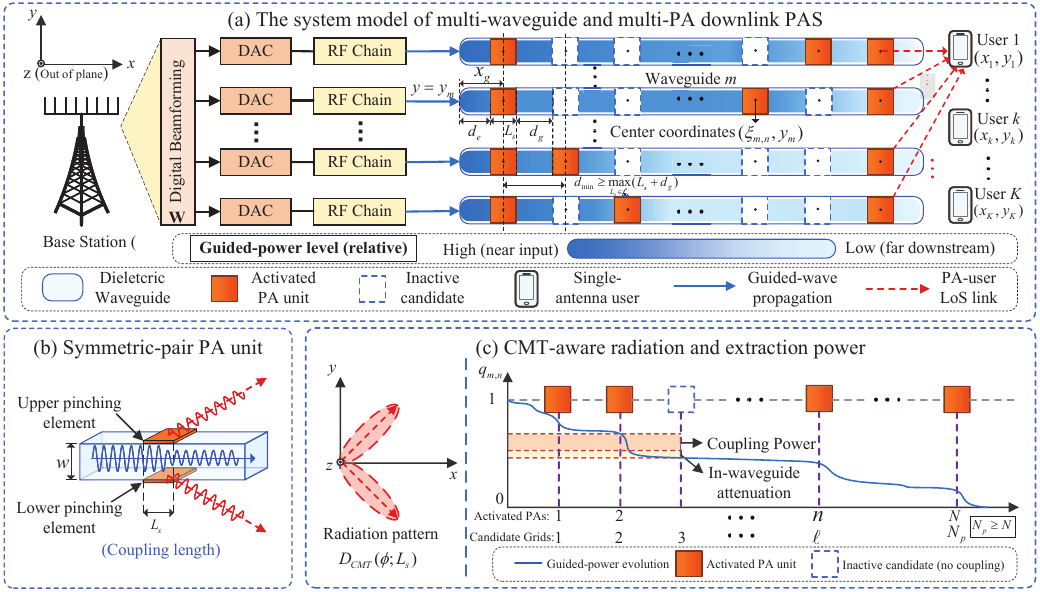}
\caption{CMT-aware multi-waveguide pinching-antenna system.}
\label{fig:system_model}\vspace{-1.5em}
\end{figure}

As shown in Fig.~\ref{fig:system_model}(a), we consider a downlink multiuser pinching-antenna system (PAS), where a base station feeds $M$ parallel dielectric waveguides to serve $K$ single-antenna users. Let $\mathcal M=\{1,\ldots,M\}$ and $\mathcal K=\{1,\ldots,K\}$ denote the waveguide and user index sets, respectively. The $M$ waveguides, each of length $L_w$, are parallel to the $x$-axis and uniformly spaced along the $y$-axis, with $y_m$ denoting the position of waveguide $m$. They are assumed sufficiently separated so that inter-waveguide coupling is negligible. Each waveguide is excited by an independently controlled feed port, and transmit beamforming is performed across the $M$ feed ports. The location of user $k$ is denoted by $\mathbf u_k=(x_k,y_k)$, $k\in\mathcal K$. Following \cite{zubair2026closed}, each effective PA unit is modeled as a symmetric pair of identical pinching elements placed at the same longitudinal coordinate on the two sides of the waveguide, as shown in Fig.~\ref{fig:system_model}(b).\footnote{The symmetric pair shares one local guided mode and is modeled as one effective PA unit with one coupling ratio and one radiation pattern.} Accordingly, each waveguide employs $N$ such effective activated PA units, indexed by $\mathcal N=\{1,\ldots,N\}$. For waveguide $m$, their ordered longitudinal center coordinates are collected in $\boldsymbol{\xi}_m =[\xi_{m,1},\ldots,\xi_{m,N}]^{\rm T}$ and $ p_{m,n}=(\xi_{m,n},y_m)$ denotes the location of PA unit $(m,n)$. For $L_s$ selected from the prescribed set $\mathcal L_s=\{L_s^{(q)}\}_{q=1}^{N_L}$, PA unit $(m,n)$ spans the
coupling section $[\xi_{m,n}-L_s/2,\xi_{m,n}+L_s/2]$. Here, the section center $\xi_{m,n}$ is used as the effective radiation and phase reference when evaluating the guided-wave phase, in-waveguide attenuation, and PA-user distance, whereas the local coordinate $\tilde{x}$ is used only in the unit-level CMT model. To assess PA-placement flexibility in the CMT-aware design, we use the unified position definition framework with $\xi_{m,n}$ continuously optimized or selected from predefined candidates for flexible and hardware-constrained deployments, respectively.

To avoid overlap for all $L_s \!\!\in\!\! \mathcal L_s$, the minimum center spacing $d_{\min}$ and end margin $x_g$ are chosen such that $d_{\min}\ge \max_{L_s\in\mathcal L_s}(L_s+d_g)$ and $x_g\ge d_e+\frac{1}{2}\max_{L_s\in\mathcal L_s}L_s$, where $d_g$ and $d_e$ are the guard distances between adjacent PA units and between a PA unit and the nearest waveguide end, respectively.
\textbf{Continuous design:} The PA centers on waveguide $m$ satisfy
\vspace{-0.6em}\begin{equation}
x_g\le \xi_{m,1}\!<\!\ldots<\xi_{m,N}\!\le\! L_w-x_g,
\xi_{m,n+1}\!-\!\xi_{m,n}\!\ge\! d_{\min},
\label{eq:continuous_position_constraints}
\vspace{-0.6em}\end{equation}
and form the continuous configuration $\boldsymbol{\Xi}_c\!\triangleq\!
\{\boldsymbol{\xi}_m\}_{m\in\mathcal M}$. Its feasible set is $\mathcal F_{\rm c}\triangleq
\{\boldsymbol{\Xi}_{\rm c}\mid
\boldsymbol{\xi}_m\ \text{satisfies
\eqref{eq:continuous_position_constraints}},
\ \forall m\in\mathcal M\}.$ \textbf{Discrete design:}
Each waveguide has $N_{\rm p}\ge N$ candidate PA positions, uniformly distributed over $[x_g,L_w-x_g]$ as
\vspace{-0.4em}\begin{equation}
\bar{x}_{\ell}=x_g+({{\ell}-1})(L_w-2x_g)/({N_p-1}),\, {\ell}=1,\ldots,N_p .
\label{eq:uniform_candidates}
\vspace{-0.4em}\end{equation}
The grid spacing satisfies $(L_w\!-\!2x_g)/(N_{\rm p}\!-\!1)\!\ge\! d_{\min}$, ensuring the minimum separation between any two activated candidates. Let $\mathbf B=[b_{m,\ell}]\in\{0,1\}^{M\times N_{\rm p}}$ denote the activation matrix, where $b_{m,\ell}=1$ indicates that candidate $\ell$ on waveguide $m$ is activated. Its feasible set is $\mathcal B\triangleq \{\mathbf B:\sum_{\ell\!=\!1}^{N_{\rm p}}b_{m,\ell}\!=\!N, \ \forall m\in\mathcal M\}$. For the activated indices $\{\ell:b_{m,\ell}=1\} =\{\ell_{m,1}<\ldots<\ell_{m,N}\}$, define $ \boldsymbol{\xi}_m(\mathbf B) = [\bar{x}_{\ell_{m,1}},\ldots,\bar{x}_{\ell_{m,N}}]^{\rm T}, \boldsymbol{\Xi}_d = \{\boldsymbol{\xi}_m(\mathbf B)\}_{m\in\mathcal M}.$ Thus, $\mathcal F_{\rm d}\triangleq \{\boldsymbol{\Xi}_d\mid\mathbf B\in\mathcal B\}$. Throughout, $\boldsymbol{\Xi}$ denotes the physical PA-position set, with $\mathcal F=\mathcal F_{\rm c}$ for continuous positioning and $\mathcal F=\mathcal F_{\rm d}$ for discrete activation.

\vspace{-1.2em}\subsection{CMT-Based Effective PA Model}
\vspace{-0.2em}The closed-form CMT model in \cite{zubair2026closed} characterizes the local response of a symmetric PA pair over a single coupling section. Assuming non-overlapping sections and negligible inter-unit reflections and mutual coupling, we apply this unit-level response independently to each effective PA unit. For PA unit $(m,n)$, let $A_{m,n}^{\rm in}$ denote the waveguide-mode amplitude incident at the entrance of its coupling section. Since no PA mode is incident at this entrance, the local boundary conditions are $A_{m,n}(0)=A_{m,n}^{\rm in}, B_{m,n}(0)=0.$ For notation brevity, the indices $(m,n)$ are omitted below, and $A_0$ denotes $A_{m,n}^{\rm in}$ for the considered PA unit. Under symmetric phase-matched coupling, the waveguide-mode amplitude $A(\tilde x)$ and the common PA-mode amplitude $B(\tilde x)$ satisfy
${dA(\tilde x)}/{d\tilde x}=-2j\kappa B(\tilde x), {dB(\tilde x)}/{d\tilde x}=-j\kappa A(\tilde x),$ where $\kappa$ is the coupling coefficient and $\tilde{x}\in[0,L_s]$ is measured from the section entrance. The resulting closed-form solution is
\vspace{-0.4em}\begin{equation}
A(\tilde x)=A_0\cos(\sqrt{2}\kappa \tilde x),\,
B(\tilde x)=-{jA_0}\sin(\sqrt{2}\kappa \tilde x)/{\sqrt{2}}.
\label{eq:cmt_closed}
\vspace{-0.4em}\end{equation}

Accounting for both symmetric pinching elements, whose normalized power is $2|B(L_s)|^2/|A_0|^2=\sin^2(\sqrt{2}\kappa L_s)$, the effective guided-power extraction ratio of one PA unit is modeled as $\rho(L_s)=\rho_{\max}\sin^2(\sqrt{2}\kappa L_s)$, where $0<\rho_{\max}\leq1$ captures nonideal coupling. Its first maximum occurs at
\vspace{-0.4em}\begin{equation}\label{eq:maxcoupling}
L_c={\pi}/{2\sqrt{2}\kappa}.
\vspace{-0.4em}\end{equation}
Although $L_s=L_c$ maximizes the local extraction ratio by the considered PA unit, it leaves the smallest residual guided-power fraction $1-\rho(L_s)$ for downstream PA units. The corresponding far-field radiation pattern is
$
F_{\rm CMT}(\phi;L_s)
=
P(\phi)F_x(\phi;L_s)F_y(\phi),
$
where $P(\phi)$, $F_x(\phi;L_s)$, and $F_y(\phi)$ denote the projection, longitudinal-radiation, and transverse-radiation factors, respectively. Their closed-form expressions are given in \cite{zubair2026closed}. The normalized in-plane directional gain is
\vspace{-0.3em}\begin{equation}
D_{\rm CMT}(\phi;L_s)
=
\frac{|F_{\rm CMT}(\phi;L_s)|^2}
{\frac{1}{2\pi}\int_0^{2\pi}|F_{\rm CMT}(\vartheta;L_s)|^2d\vartheta}.
\label{eq:D_cmt}
\vspace{-0.3em}\end{equation}
Thus, the guided-power extraction ratio $\rho(L_s)$ and the normalized directional gain $D_{\rm CMT}(\phi;L_s)$ characterize the normalized response shared by all identical PA units, whereas their PA-dependent incident powers and PA-user geometries are incorporated into the multi-PA channel below.
\vspace{-1.3em}\subsection{CMT-Aware Multi-PA Channel}\label{Loss-Aware Multi-PA Channel}
Using the unit-level responses above, we construct the multi-PA channel by incorporating sequential power extraction, guided-wave attenuation, and PA-user propagation. Under the coplanar LoS geometry
of the two-dimensional CMT model, the distance and departure angle from PA unit
$(m,n)$ to user $k$ are $r_{k,m,n}\!=\!\|\mathbf u_k-\mathbf p_{m,n}\|_2$ and $\phi_{k,m,n}\!=\!\operatorname{atan2} (y_k\!-y_m,x_k\!-\xi_{m,n})$, respectively. With each waveguide fed at $x\!=0$, let $\alpha_g$ denote the guided-power attenuation coefficient in dB/m. Since the PA units are ordered along each waveguide and sequentially excited by the same guided wave, the normalized incident guided-power factor at unit $(m,n)$ is modeled as
\vspace{-0.4em}\begin{equation}
q_{m,n}
=
10^{-\alpha_g \xi_{m,n}/10}
\prod\nolimits_{i=1}^{n-1}
\left[1-\rho(L_s)\right],
\label{eq:q}
\vspace{-0.4em}\end{equation}
where the product accounts for power extraction by the upstream PA units and equals one for $n=1$. The useful radiated-power factor is $ p_{m,n}^{\rm rad} = \eta_r\rho(L_s)q_{m,n},$ where $0<\eta_r\le1$ denotes the radiation efficiency. Thus, increasing $\rho(L_s)$ strengthens the radiation of the current PA unit but reduces the guided power available to downstream units. Defining $\mathbf h_k$ waveguide-domain channel vector of user $k$, the channel of PA unit $(m,n)$ between waveguide $m$ and user $k$ is
\vspace{-0.4em}\begin{equation}
\begin{aligned}
{\tilde{g}}_{k,m,n}
&=
\frac{\lambda}{4\pi r_{k,m,n}}
\sqrt{p_{m,n}^{\rm rad}D_{\rm CMT}(\phi_{k,m,n};L_s)} \\
&\quad \times
e^{-j(\beta_g \xi_{m,n}+k_0r_{k,m,n})}.
\end{aligned}
\label{eq:g_kmn}
\vspace{-0.4em}\end{equation}
Here, $D_{\rm CMT}$ serves as a normalized in-plane power weighting under a common reference gain, rather than an absolute three-dimensional antenna gain. Moreover, $\lambda$ is the free-space wavelength, $k_0=2\pi/\lambda$ is the free-space wavenumber, and $\beta_g$ is the guided-wave propagation constant. The two phase terms represent guided-wave and free-space propagation, respectively. The fields from the $N$ PA units on waveguide $m$ combine coherently at user $k$, yielding $h_{k,m}=\sum_{n=1}^{N}\widetilde{g}_{k,m,n}$ and $\mathbf h_k(\boldsymbol{\Xi},L_s) =[h_{k,1},\ldots,h_{k,M}]^{\rm T}$.

\vspace{-1.2em}\subsection{Received Signal and Problem Formulation}
\vspace{-0.2em}Let $\mathbf W=[\mathbf w_1,\ldots,\mathbf w_K]\in\mathbb C^{M\times K}$ denote the feed-port beamforming matrix, and let $\mathbf s=[s_1,\ldots,s_K]^{\rm T}$ denote the information-symbol vector with $\mathbb E\{\mathbf s\mathbf s^{\rm H}\}=\mathbf I_K$. The baseband signal injected into the $M$ waveguide feeds is
$ \mathbf x = \mathbf W\mathbf s = \sum\nolimits_{k=1}^{K}\mathbf w_k s_k,$ where $\mathbf w_k\in\mathbb C^{M\times1}$ is the beamforming vector for user $k$. The total power fed into the waveguides is
$P_{\rm tx}=\|\mathbf W\|_{\rm F}^{2}=\sum\nolimits_{k=1}^{K}\|\mathbf w_k\|_2^2$. Given the channel vector
$\mathbf h_k(\boldsymbol{\Xi},L_s)$, the received signal at user $k$ is
\vspace{-0.4em}\begin{equation}
y_k
=
\mathbf h_k(\boldsymbol{\Xi},L_s)^{\rm H}\mathbf w_k s_k
+
\sum\nolimits_{j\ne k}\mathbf h_k(\boldsymbol{\Xi},L_s)^{\rm H}\mathbf w_j s_j
+
n_k,
\label{eq:received_signal}
\vspace{-0.4em}\end{equation}
where $n_k\sim\mathcal{CN}(0,\sigma_k^2)$. Therefore, the SINR of user $k$ is
\vspace{-0.4em}\begin{equation}
\gamma_k(\mathbf W,\boldsymbol{\Xi}, L_s)
=
\frac{|\mathbf h_k(\boldsymbol{\Xi},L_s)^{\rm H}\mathbf w_k|^2}
{\sum_{j\ne k}|\mathbf h_k(\boldsymbol{\Xi},L_s)^{\rm H}\mathbf w_j|^2+\sigma_k^2}.
\label{eq:sinr}
\vspace{-0.4em}\end{equation}

For implementation consistency, all effective PA units use a common coupling length $L_s\in\mathcal L_s$. Since $L_s$ jointly determines the power-extraction ratio and directional radiation pattern, it is optimized together with the beamforming and PA positions to minimize the transmit power subject to the user-specific SINR requirements. The resulting problem is formulated as
\vspace{-0.4em}\begin{equation}
\begin{aligned}
\mathcal P_0:\quad
\min_{\mathbf W,\,\boldsymbol{\Xi},\,L_s}
& \|\mathbf W\|_{\rm F}^{2}\\
{\rm s.t.}
& \gamma_k(\mathbf W,\boldsymbol{\Xi},L_s)
\ge \Gamma_k,
\; \forall k\in\mathcal K,\\
& L_s\in\mathcal L_s,
\boldsymbol{\Xi}\in\mathcal F,
\end{aligned}
\label{eq:P0}
\vspace{-0.4em}\end{equation}
where $\Gamma_k$ is the target SINR of user $k$, $\mathcal L_s$ is
the prescribed finite coupling-length set, and $\mathcal F$ is defined in Section \ref{System_Geometry}.

\vspace{-0.8em}\section{CMT-Aware PA and Beamforming Design}
To seek a high-quality feasible solution to $\mathcal P_0$, we use a two-level framework. The outer loop enumerates $L_s^{(q)}\in\mathcal L_s$, while the inner loop performs element-wise continuous-position refinement or BPSO-based discrete activation search. The ZF transmit-power metric $P_{\rm ZF}$ ranks the trial configurations without repeated beamforming solves. For each coupling length, the top-ranked configuration is evaluated by solving the corresponding fixed-configuration beamforming problem, with the minimum-power feasible configuration pair selected.
\vspace{-1em}\subsection{Fixed-Position Beamforming}
\vspace{-0.2em}For given $(\boldsymbol{\Xi},L_s)$, the user channels are fixed. Since each beamforming vector can be independently phase-rotated without affecting the transmit power or any user SINR, $\mathbf h_k^{\rm H}\mathbf w_k$ can be taken as real and nonnegative without loss of optimality. Let $\mathbf W_{-k}$ denote $\mathbf W$ with its $k$-th column removed. The fixed-position beamforming subproblem is then equivalently cast as
\begin{equation}
\begin{aligned}
\min_{\mathbf W}\,
& \|\mathbf W\|_{\rm F}^{2} \\
{\rm s.t.}\,
& \operatorname{Im}\{\mathbf h_k^{\rm H}\mathbf w_k\}=0,\; \operatorname{Re}\{\mathbf h_k^{\rm H}\mathbf w_k\}\ge 0, \\
&\left\|\left[\mathbf h_k^{\rm H}\mathbf W_{-k},\,\sigma_k\right] \right\|_2\le{\operatorname{Re}\{\mathbf h_k^{\rm H}\mathbf w_k\}}{ /\sqrt{\Gamma_k}},\forall k\in\mathcal K .
\end{aligned}
\label{eq:PB_socp}
\end{equation}
Let $\mathbf W^{\star}(\boldsymbol{\Xi},L_s)$ denote the optimal solution of \eqref{eq:PB_socp}. The corresponding minimum transmit power is $ P(\boldsymbol{\Xi},L_s) \triangleq \left\| \mathbf W^{\star}(\boldsymbol{\Xi},L_s) \right\|_{\rm F}^{2}$ used for final candidate evaluation. Since solving \eqref{eq:PB_socp} for every inner-loop trial is computationally costly, a closed-form ZF power metric is derived for candidate ranking.

\vspace{-1em}\subsection{ZF-Based Configuration Ranking}
\vspace{-0.2em}To reduce the inner-loop complexity, we use a closed-form ZF transmit-power metric to rank the trial PA positions. This metric is used only for candidate comparison, whereas $P(\boldsymbol{\Xi},L_s)$ is evaluated for the configuration returned by the inner search. For given $(\boldsymbol{\Xi},L_s)$, define the effective channel matrix as $\mathbf H(\boldsymbol{\Xi},L_s)\!=\![\mathbf h_1,\ldots,\mathbf h_K]^{\rm H} \in\mathbb C^{K\times M}.$ For $\operatorname{rank}(\mathbf H)=K$, define $\mathbf D\triangleq\operatorname{diag}\!\left( \sqrt{\Gamma_1}\sigma_1,\ldots,\sqrt{\Gamma_K}\sigma_K \right)$. The minimum-norm ZF beamformer satisfying $\mathbf H\mathbf W_{\rm ZF}=\mathbf D$ is
\vspace{-0.4em}\begin{equation}
\mathbf W_{\rm ZF}
=
\mathbf H^{\rm H}
(\mathbf H\mathbf H^{\rm H})^{-1}
\mathbf D.
\label{eq:zf_beamformer}
\vspace{-0.4em}\end{equation}

Further define $\boldsymbol{\Lambda}\triangleq\mathbf D\mathbf D^{\rm H}=\operatorname{diag}
\left(\Gamma_1\sigma_1^2,\ldots,\Gamma_K\sigma_K^2\right).$ The corresponding ZF-based transmit-power metric is
\vspace{-0.3em}\begin{equation}
\begin{aligned}
P_{\rm ZF}(\boldsymbol{\Xi},L_s)
\triangleq
\|\mathbf W_{\rm ZF}\|_{\rm F}^{2}
=
\operatorname{tr}
\left[
(\mathbf H\mathbf H^{\rm H})^{-1}
\boldsymbol{\Lambda}
\right].
\end{aligned}
\label{eq:zf_metric}
\vspace{-0.3em}\end{equation}
For rank-deficient trials, we set $P_{\rm ZF}=+\infty$, excluding them whenever a full-row-rank candidate exists. For full-row-rank $\mathbf H$, $\mathbf W_{\rm ZF}$ nulls inter-user interference and satisfies the prescribed SINR targets, making $P_{\rm ZF}$ a feasible upper bound on the minimum fixed-configuration transmit power. Its evaluation requires only constructing the CMT-aware channel and inverting a $K\times K$ matrix, avoiding an SOCP solve for each trial. Thus, $P_{\rm ZF}$ provides a low-complexity ranking metric for both continuous positioning and discrete activation.

\vspace{-1.5em}\subsection{Continuous and Discrete PA Optimization}
\vspace{-0.4em}\textbf{Continuous design:} For each fixed $L_s^{(q)}$, the PA coordinates are refined sequentially while all other coordinates held fixed. When updating $\xi_{m,n}$, the boundary and spacing
constraints define the feasible interval $\mathcal I_{m,n}=[\xi_{m,n}^{\min},\xi_{m,n}^{\max}]$, where
\vspace{-0.6em}\begin{equation}
\begin{aligned}
\xi_{m,n}^{\min}
&=
\begin{cases}
x_g, & n=1,\\
\xi_{m,n-1}+d_{\min}, & n>1,
\end{cases}
\\
\xi_{m,n}^{\max}
&=
\begin{cases}
L_w-x_g, & n=N,\\
\xi_{m,n+1}-d_{\min}, & n<N.
\end{cases}
\end{aligned}
\label{eq:coordinate_bounds}
\vspace{-0.6em}\end{equation}
\vspace{-0.2em}A finite trial set is constructed as $\mathcal G_{m,n} \triangleq\{\zeta_{m,n}^{(g)}\}_{g=1}^{G}\subseteq\mathcal I_{m,n}, $ where $\zeta_{m,n}^{(g)}$ denotes the $g$-th trial position. For any $\zeta\in\mathcal G_{m,n}$, let
$\boldsymbol{\Xi}_{m,n}(\zeta)$ denote the PA-position set with
$\xi_{m,n}=\zeta$ and all other coordinates unchanged. The coordinate is updated as
\vspace{-0.5em}\begin{equation}
\xi_{m,n}
\leftarrow
\arg\min\nolimits_{\zeta\in\mathcal G_{m,n}}
P_{\rm ZF}
(
\boldsymbol{\Xi}_{m,n}(\zeta),L_s^{(q)}
).
\label{eq:continuous_coordinate_update}
\vspace{-0.5em}\end{equation}
For each trial position, the complete CMT-aware channel is recomputed to account for the resulting changes in guided-wave attenuation, propagation distance and phase, and directional radiation gain. The sequential updates over all $MN$ coordinates are repeated until convergence or the maximum number of sweeps is reached, yielding  $N_{\rm cand}$ smallest-ZF-power configurations $\{\widehat{\boldsymbol{\Xi}}_{\rm c}^{(q,\nu)}\}_{\nu}^{N_{cand}}$ for all $q=1,\ldots,N_L$.

\textbf{Discrete design:}
For each $L_s^{(q)}\in\mathcal L_s$, the activation pattern is optimized via BPSO. At iteration $t$, particle $\iota=1,\ldots,N_{\rm pop}$ represents an activation matrix $\mathbf B^{(\iota,t)}$. Its velocity is updated using the inertia, personal-best, and global-best components and then mapped through the sigmoid function to an activation score for each candidate position. This block-wise top-$N$ selection satisfies the fixed-cardinality constraint, while the prescribed candidate spacing ensures the minimum PA separation. Each activation pattern is mapped to the corresponding PA positions $\boldsymbol{\Xi}(\mathbf B^{(\iota,t)})$. The CMT-aware channel is then recomputed, and the particle fitness is evaluated as $P_{\rm ZF}(\boldsymbol{\Xi}\bigl(\mathbf B^{(\iota,t)}\bigr), L_s^{(q)} ).$ The personal-best and global-best activation patterns are updated accordingly. The iterations continue until convergence or the maximum number $I_{\rm pso}$ is reached, yielding $N_{\rm cand}$ smallest-ZF-power configurations $\{\widehat{\mathbf B}^{(q,\nu)}\}_{\nu=1}^{N_{\rm cand}}$ with $ \widehat{\boldsymbol{\Xi}}_{\rm d}^{(q,\nu)} \triangleq \boldsymbol{\Xi} ( \widehat{\mathbf B}_{\rm gbest}^{(q,\nu)})$ for all $q=1,\ldots,N_L$.

\vspace{-0.3em}Overall, for a unified notation, let $\widehat{\boldsymbol{\Xi}}^{(q,\nu)} =\widehat{\boldsymbol{\Xi}}_{\rm c}^{(q,\nu)}$ for continuous positioning and $\widehat{\boldsymbol{\Xi}}^{(q,\nu)} =\widehat{\boldsymbol{\Xi}}_{\rm d}^{(q,\nu)}$ for discrete activation. For each $(\widehat{\boldsymbol{\Xi}}^{(q,\nu)},L_s^{(q,\nu)})$, the fixed-position beamforming subproblem is solved to obtain $P^{(q,\nu)} = P\!( \widehat{\boldsymbol{\Xi}}^{(q,\nu)},L_s^{(q)}) $ and the corresponding beamformer $\mathbf W^{(q,\nu)}$. Let $\mathcal {F}_{\rm{cand}}$ denote the feasible $(q,\nu)$. The final candidate is selected as
\vspace{-0.6em}\begin{equation}
(q^\star,\nu^\star)
\in
\arg\min\nolimits_{(q,\nu)\in\mathcal F_{\rm {cand}}} P^{(q,\nu)},
\label{eq:selected_candidate}
\vspace{-0.6em}\end{equation}
yielding $ L_s^\star=L_s^{(q^\star)}, {\boldsymbol{\Xi}}^\star=\widehat{\boldsymbol{\Xi}}^{(q^\star,\nu^\star)}, \mathbf W^\star=\mathbf W^{(q^\star,\nu^\star)}. $ The overall procedure is summarized in Algorithm~\ref{alg:proposed}.

Let $I_{\rm c}$ denote the number of continuous refinement sweeps, $N_{\rm pop}$ the BPSO population size, $I_{\rm pso}$ the number of BPSO iterations, and $N_{\rm cand}$ the number of top-ranked
configurations retained for SOCP evaluation at each coupling length. Moreover, let $\mathcal C_{\rm ZF}$ and $\mathcal C_{\rm SOCP}$ denote the complexities of one ZF-metric evaluation and one SOCP solve, respectively. The continuous and discrete designs have complexities $ \mathcal O \left( N_L I_{\rm c}MNG\mathcal C_{\rm ZF}
+ N_L N_{\rm cand}\mathcal C_{\rm SOCP} \right)$ and $ \mathcal O \left(N_LN_{\rm pop}I_{\rm pso}\mathcal C_{\rm ZF} + N_L N_{\rm cand}\mathcal C_{\rm SOCP} \right),$ respectively.

\begin{algorithm}[!t]
\caption{Proposed CMT-Aware PASS Optimization}
\label{alg:proposed}
\begin{algorithmic}[1]
\STATE \textbf{Input:}
$\mathcal L_s$, $N_{\rm cand}$, CMT parameters, and design mode.
\FOR{$q=1,\ldots,N_L$}
    \IF{continuous positioning}
        \STATE Retain the top $N_{\rm cand}$ configurations
        $\{\widehat{\boldsymbol{\Xi}}^{(q,\nu)}\}_{\nu=1}^{N_{\rm cand}}$
        from the $P_{\rm ZF}$-based coordinate search in
        \eqref{eq:continuous_coordinate_update}.
    \ELSE
        \STATE Retain the top $N_{\rm cand}$ BPSO configurations
        $\{\widehat{\boldsymbol{\Xi}}^{(q,\nu)}\}_{\nu=1}^{N_{\rm cand}}$
        using $P_{\rm ZF}$ as the fitness metric.
    \ENDIF
    \FOR{$\nu=1,\ldots,N_{\rm cand}$}
        \STATE Solve \eqref{eq:PB_socp} at
        $(\widehat{\boldsymbol{\Xi}}^{(q,\nu)},L_s^{(q)})$
        for $(P^{(q,\nu)},\mathbf W^{(q,\nu)})$.
    \ENDFOR
\ENDFOR
\STATE Select
$(q^\star,\nu^\star)\in
\arg\min_{(q,\nu)\in\mathcal F_{\rm{cand}}}P^{(q,\nu)}$.
\STATE \textbf{return}
$(L_s^\star,\boldsymbol{\Xi}^\star,\mathbf W^\star)$.
\end{algorithmic}
\end{algorithm}

\begin{figure*}[!t]
\centering
\subfloat[]{%
\includegraphics[width=0.24\textwidth]{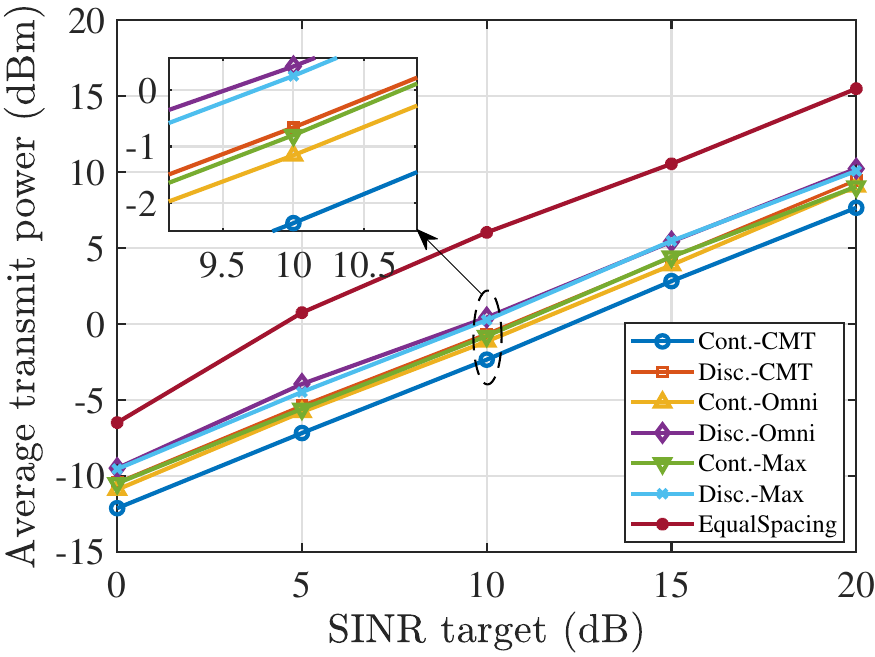}
\label{fig:sinr_sweep}}
\hfill
\subfloat[]{%
\includegraphics[width=0.24\textwidth]{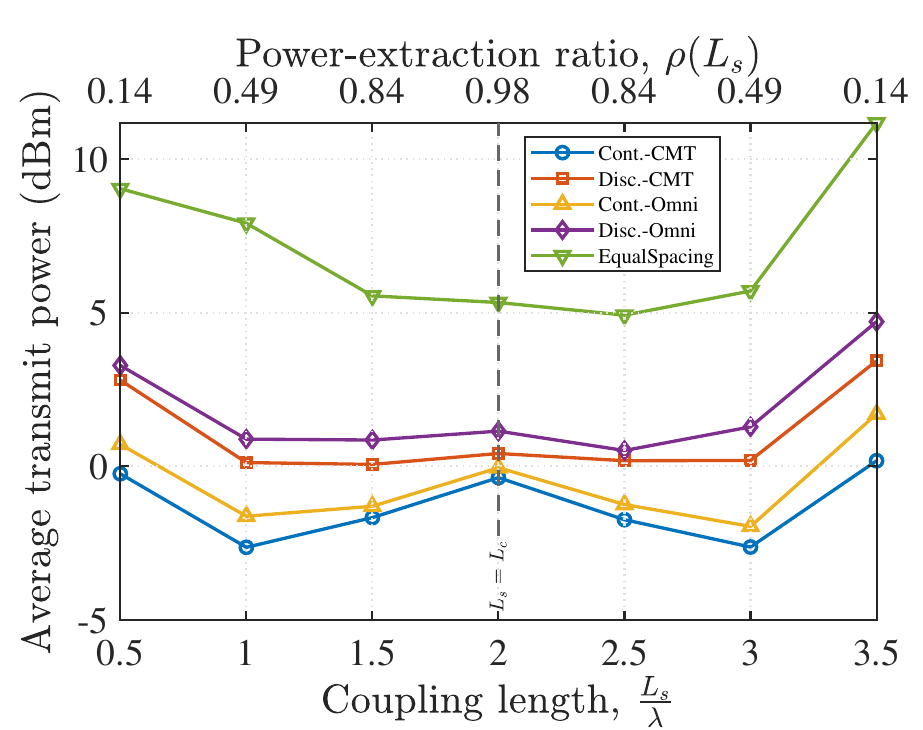}
\label{fig:coupling_sweep}}
\hfill
\subfloat[]{%
\includegraphics[width=0.24\textwidth]{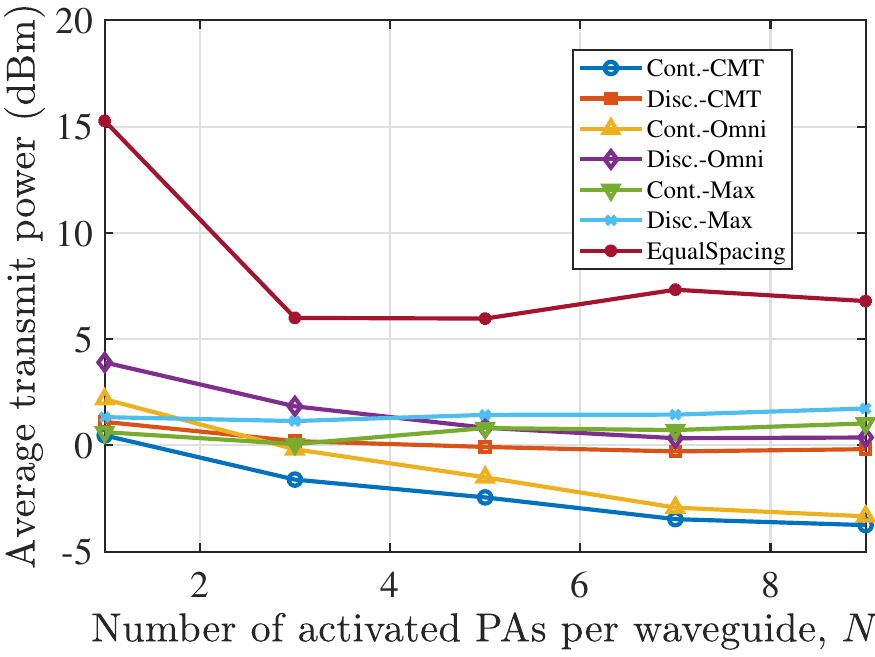}
\label{fig:num_pa_sweep}}
\hfill
\subfloat[]{%
\includegraphics[width=0.24\textwidth]{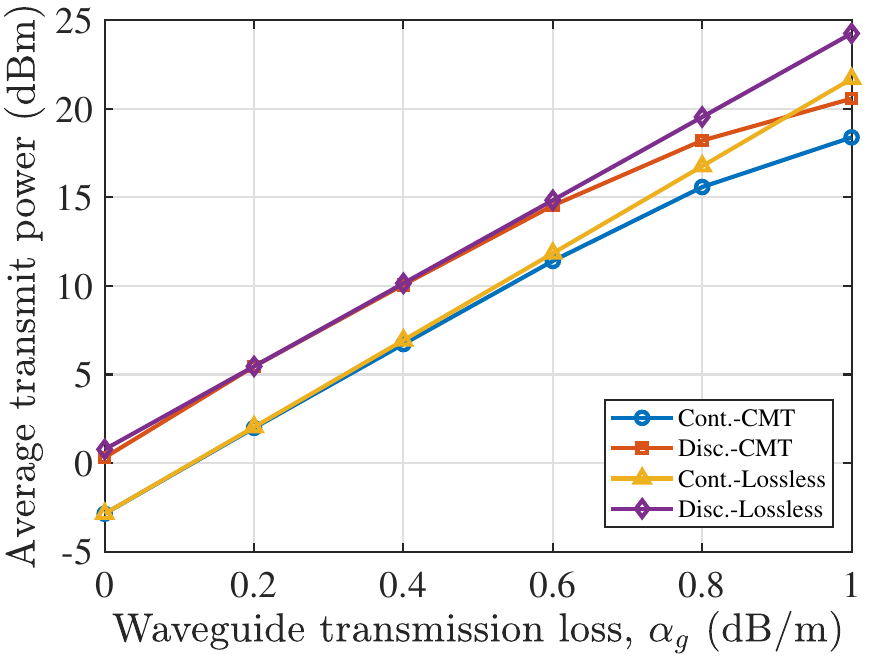}
\label{fig:loss_sweep}}

\caption{Average transmit power of different schemes.
(a) Average transmit power versus SINR target.
(b) Average transmit power versus normalized coupling length $L_s/\lambda$.
(c) Average transmit power versus the number of activated PAs.
(d) Average transmit power versus waveguide transmission loss coefficient.}\vspace{-2em}
\label{fig:performance_comparison}
\end{figure*}

\vspace{-1em}\section{Numerical Results}
\vspace{-0.5em}Unless otherwise stated, the simulation parameters follow \cite{zubair2026closed,xu2026power}. We consider a $60$-GHz downlink PAS with $M=4$ parallel waveguides serving $K=3$ users in a $10~{\rm m}\times6.75~{\rm m}$ region. Each waveguide has length $L_w=10$ m and supports $N=3$ active PAs. The waveguides are uniformly spaced along the $y$-axis over $[-0.75,0.75]$ m, while the users are independently and uniformly distributed in $[0.5,9.5]\times[2,6]~{\rm m}^2$. For the discrete design, each waveguide contains $N_{\rm p}=21$ uniformly spaced candidate positions. The radiation efficiency, maximum coupling ratio, and guided-wave attenuation coefficient are set to $\eta_r=0.8$, $\rho_{\max}=0.98$, and $\alpha_g=0.15$ dB/m, respectively. We set $L_c=2\lambda$ by choosing $\kappa=\pi/(2\sqrt{2}L_c)$ and search over $L_s/\lambda\in\{0.5,1,1.5,2,2.5,3,3.5\}$. To separate the effects of PA-position optimization and CMT-aware channel modeling, we consider several benchmark schemes.  \textbf{Cont.-CMT} and \textbf{Disc.-CMT} denote the proposed CMT-aware continuous-position and discrete-activation designs, respectively. \textbf{Cont.-Omni} and \textbf{Disc.-Omni} optimize the PA configurations using an omnidirectional model but are evaluated under the CMT-aware directional channel. This reveals the performance loss caused by radiation-model mismatch. \textbf{Cont.-Max} and \textbf{Disc.-Max} fix $L_s=L_c$, whereas \textbf{EqualSpacing} uniformly places the active PAs and optimizes only the transmit beamforming. For the attenuation study, \textbf{Cont.-Lossless} and \textbf{Disc.-Lossless} optimize the PA design without attenuation and are evaluated under the loss-aware CMT channel.


Fig.~\ref{fig:performance_comparison}\subref{fig:sinr_sweep} shows the average transmit power versus the minimum SINR requirement. As expected, the required power increases with the SINR target. Under both design modes, the CMT-aware schemes outperform their omnidirectional counterparts. By ignoring the coupling-dependent directional gain, the omnidirectional model may select PA positions or activation patterns that weaken the desired links or make spatial interference suppression less effective, thereby increasing the beamforming power required to meet the same SINR targets. \textbf{Cont.-CMT} performs best owing to its continuous positioning flexibility, whereas \textbf{Disc.-CMT} incurs a moderate codebook loss but remains superior to \textbf{EqualSpacing}. Moreover, \textbf{Cont.-Max} and \textbf{Disc.-Max} cannot fully
exploit the multi-PA aperture because maximum upstream extraction reduces downstream guided power and weakens the resulting channel contributions.

Fig.~\ref{fig:performance_comparison}\subref{fig:coupling_sweep} illustrates the impact of the normalized coupling length $L_s/\lambda$, with the upper axis indicating the corresponding per-PA extraction ratio $\rho(L_s)$. Although $\rho(L_s)$ reaches its first maximum at $L_s=L_c=2\lambda$, this point does not minimize the system transmit power. This is because $L_s$ jointly determines power extraction ratio, radiation directivity, and the residual guided power available to downstream PAs. A large $\rho(L_s)$ strengthens the upstream PA contributions but leaves less residual power for downstream PAs, thereby degrading the effective multiuser channel. Hence, the power-minimizing $L_s$ balances local power extraction against the aggregate channel contribution of all active PAs. The gains of the CMT-aware designs further confirm that optimizing $L_s$ based only on $\rho(L_s)$ or an omnidirectional model is insufficient. Its coupling-dependent radiation pattern must be jointly considered with the PA configuration and transmit beamforming.


Fig.~\ref{fig:performance_comparison}\subref{fig:num_pa_sweep} evaluates the number $N$ of active PAs per waveguide. Increasing $N$ generally lowers the transmit power by enlarging the distributed aperture and providing more flexibility for shaping the user channels. However, the gain gradually saturates because upstream activated PAs progressively deplete the guided power available to downstream units, thereby limiting effective aperture utilization. \textbf{Cont.-CMT} benefits most from the increased positioning freedom, while \textbf{Disc.-CMT} also outperforms \textbf{EqualSpacing} despite using a discrete candidate grid. The
poorer performance of \textbf{Cont.-Max} and \textbf{Disc.-Max} at large $N$ further shows that additional PAs are beneficial only when their positions or activation patterns account for the coupling-length-dependent power evolution.


Fig.~\ref{fig:performance_comparison}\subref{fig:loss_sweep} examines the impact of $\alpha_g$ in a long-waveguide scenario. The required power increases with $\alpha_g$ because the accumulated in-waveguide attenuation reduces the guided power reaching downstream PAs. Moreover, this loss depends on the PA position and therefore changes the relative channel contributions of the active PAs, rather than merely scaling all channels by the same factor. Consequently, a lossless model overestimates downstream radiation and may select unsuitable PA positions or activation patterns. The widening gap from the lossless baselines confirms the importance of accounting for guided-wave attenuation in long-waveguide PAS deployments.

\vspace{-0.8em}\section{Conclusion}
\vspace{-0.2em}This letter investigated CMT-aware transmit-power minimization for multiuser PAS. A closed-form CMT radiation model was incorporated into the multi-PA channel to capture coupling-dependent power extraction and directivity, guided-power depletion, in-waveguide attenuation, and coherent multi-PA field superposition. Based on this model, element-wise continuous positioning and BPSO-based discrete activation were developed. The results showed that CMT characteristics significantly affect PA positioning, activation, and beamforming. In particular, maximum local coupling is not system-optimal because excessive upstream extraction weakens downstream PA contributions, while the gain from additional PAs eventually saturates as guided power is depleted. The guided-wave attenuation also becomes increasingly important for long waveguides, and finite-codebook activation provides a favorable performance-complexity tradeoff. Future work will consider three-dimensional near-field PAS modeling.

\balance

\end{document}